\documentclass[prl,epsfig,amsmath,amssymb,twocolumn]{revtex4}
\usepackage{graphicx}
\usepackage{dcolumn}
\usepackage{bm}

\begin{document}

\title{Bias-dependent Contact Resistance in Rubrene Single-Crystal Field-Effect Transistors}
\author{Anna Molinari\footnote{a.molinari@tnw.tudelft.nl}, Ignacio Guti\'{e}rrez, Iulian N. Hulea, Saverio Russo and Alberto F.  Morpurgo }
\affiliation{Kavli Institute of Nanoscience, Delft University of
Technology, Lorentzweg 1, 2628CJ Delft, The Netherlands}

\begin{abstract}
We report a systematic study of the bias-dependent contact
resistance in rubrene single-crystal field-effect transistors with
Ni, Co, Cu, Au, and Pt electrodes. We show that the reproducibility
in the values of contact resistance strongly depends on the metal,
ranging from a factor of two for Ni to more than three orders of
magnitude for Au. Surprisingly, FETs with Ni, Co, and Cu contacts
exhibits an unexpected reproducibility of the bias-dependent
differential conductance of the contacts, once this has been
normalized to the value measured at zero bias. This reproducibility
may enable the study of microscopic carrier injection processes into
organic semiconductors.
\end{abstract}

\maketitle

Improvements in the material control of organic thin films
occurred during the past decade are enabling the reproducible,
low-cost fabrication of organic field-effect transistors (FETs)
with mobility values in the range 0.1 - 1 cm$^2$/Vs \cite{Halik}.
These values are sufficient for the development of new
applications in the field of plastic electronics \cite{Gelinck}.
However, the fabrication of high-quality electrical contacts for
organic transistors has not witnessed comparable progress
\cite{Pesavento et al.}, and low-quality contacts are now starting
to pose limits to the performance of organic FETs. Specifically,
with mobility values in between 0.1 and 1 cm$^2$/Vs, the contact
resistance -typically larger than 1 kOhm cm even in the best
devices- limits the transistor performance as soon as the channel
length becomes smaller than $\simeq 10 \mu$m \cite{Klauk},
preventing the possibility of device downscaling. The
irreproducibility of the contact resistance makes the situation
even worse: for gold-contacted pentacene thin-film FETs, for
instance, the spread in contact resistance values was recently
observed to exceed three orders of magnitude (from 2 kOhm cm to
more than 1 MOhm cm) \cite{Meijer}. The current lack of
understanding of the microscopic carrier injection
\cite{Gershenson} processes from a metal electrode into an organic
semiconductor does not help us to determine the causes of the
observed irreproducibility and more systematic experiments are
needed to explore the performance of different
metals as contact materials.\\
Here we report systematic transport measurements of rubrene
($C_{42}H_{28}$) single crystal FETs with electrodes made of five
different metals (Ni, Co, Cu, Au, Pt). All the transistors have
been fabricated with a sufficiently short channel length, so that
the total device resistance is entirely dominated by the contacts.
By studying more than 250 contact-dominated devices we have
collected enough statistics to determine the average contact
resistance, its spread in values, as well as its bias dependence.
We find significant differences between the different metals. In
particular, for Nickel -which exhibits the lowest resistance- the
spread is only a factor of two, for Cobalt and Copper slightly
more than one order of magnitude, and for Gold more than three
orders of magnitude (Platinum seems to behave similarly to Gold,
but the number of devices tested was not sufficient to make more
quantitative statements). We also find that for Ni, Cu, and Co
(but nor for Au and Pt) the bias dependence of the contact
resistance normalized to the resistance measured at zero-bias
exhibits an excellent reproducibility, and can be interpreted in
terms of two back-to-back Schottky diodes connected in series. Our
results suggest that organic single-crystal FETs with Ni, Co, and
Cu contacts may be suitable for the investigation of the
microscopic carrier injection
processes at a metal/organic interface.\\
\begin{figure}[h]
\centering
\includegraphics[width=0.8\columnwidth]{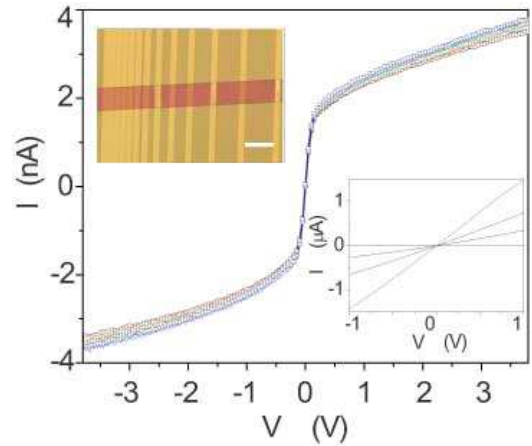}
\noindent{\caption{$I-V$ curves for a contact dominated FET ($L=5
\mu$m for 4 different gate voltages, $V_{G}=-20,-30,-40,-50 \ V$)
showing non-linear behavior, different from the usual FET
characteristic reported in the inset (shown in the lower inset from
measurements on long channel device). Top left inset: optical
microscope image of one of our devices (the white bar is $50\mu$m
long).} \label{Fig1}}
\end{figure}
Some of the metals (i.e., Cu and Co) investigated here have not been
used previously for the fabrication of organic FETs. In fact, the
vast majority of past experiments have relied on noble metal
electrodes, especially gold, whose choice is motivated by the high
value of their work-function and by the stability against oxidation
in air. These criteria, although plausible, have never been
thoroughly investigated. Our recent and unexpected finding of record
low contact resistance in devices with Nickel electrodes (100
$\Omega$cm) \cite{Iulian1} clearly underscores the importance to
explore a broader class of materials.

The rubrene single crystal FETs are fabricated by lamination of thin
($\approx 1 \mu$m thick), free-standing crystals grown by vapor
phase transport, onto a highly doped Si substrate, covered by a 200
nm thick thermally grown SiO$_2$ layer with pre-defined metal
electrodes (see Ref. \cite{Ruth1}). The contacts are fabricated by
means of electron-beam lithography, evaporation and lift-off and
their geometry is chosen so that many FETs with channel length
varying from 200 nm to 50 $\mu$m can be fabricated on the same
crystal (see inset in Fig. 1). Our earlier studies have shown that
the room-temperature carrier mobility in rubrene single-crystal FETs
with SiO$_2$ gate dielectric is narrowly spread around 4 cm$^2$/Vs
\cite{Arno and Iul}, which allows us to estimate the maximum channel
length $L$ for which the channel resistance can be neglected with
respect to the contact resistance. Specifically, for FETs with
Nickel contacts (whose resistance is normally lower than 1 kOhm cm)
we have confined our measurements to devices with $L$ smaller than 2
$\mu$m. For devices with electrodes made of the other materials,
where the contact resistance is higher, devices with channel lengths
up to $\simeq 20 \mu$m have also been used. With the channel
resistance being negligible in all cases, the device resistance
corresponds to the total contact resistance
(i.e., to the sum of the source and drain resistances).\\
Fig. 1 shows the $I-V$ characteristics of a Cu-contacted FET with
$L=5 \mu$m, measured \cite{Measurements} at different values of
the gate voltage (much larger than the threshold voltage). The
curves are essentially independent of the gate voltage and exhibit
a very pronounced non-linear increase of the source-drain current
at low bias. These $I-V$ characteristics are markedly different
from those of conventional transistors where the resistance is
dominated by the channel, as illustrated in the inset of Fig. 1. A
non-linear and gate-voltage independent $I-V$ curve similar to
that shown in Fig. 1 has been observed in short-channel devices
fabricated with all the different metal electrodes, and to analyze
this behavior in more detail we look at the differential
conductance $dI/dV$ of the devices (obtained by numerical
differentiation of the $I-V$ curves; see Fig. 2a). In the
differential conductance plots, the non-linearity present in the
$I-V$ characteristics produces a narrow peak around zero bias on a
voltage scale comparable to $kT/e$, with the precise value
being different for the different metals.\\
\begin{figure}[h]
\centering
\includegraphics[width=0.8\columnwidth]{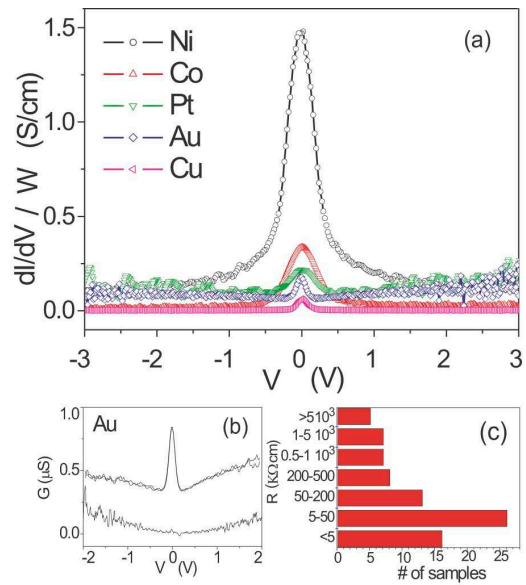}
\noindent{\caption{(a) Differential conductance normalized to the
crystal width ($W$) measured on FETs with different metal
electrodes. (b) Differential conductance for two different
short-channel gold-contacted FETs. (c) Histogram showing the
spread in contact resistance for Au electrodes.} \label{Fig2}}
\end{figure}
In order to compare the contact properties of FETs contacted with
the five different metals we have carefully analyzed the
reproducibility of the measured differential resistance. We find
that the level of reproducibility depends on the specific metal
used. The differential conductance measured on two different
short-channel gold-contacted FETs is shown in Fig. 2b, which
illustrates the poor reproducibility of these devices. In fact,
for Au contacts (and similarly for Pt) the bias dependence of the
differential conductance exhibits large differences in different
samples and a zero-bias peak is observed only in a few devices. At
the same time, the absolute value of the resistance measured at
low bias is spread over three orders of magnitude (see Fig. 2c).\\
\begin{figure}[h]
\centering
\includegraphics[width=0.85\columnwidth]{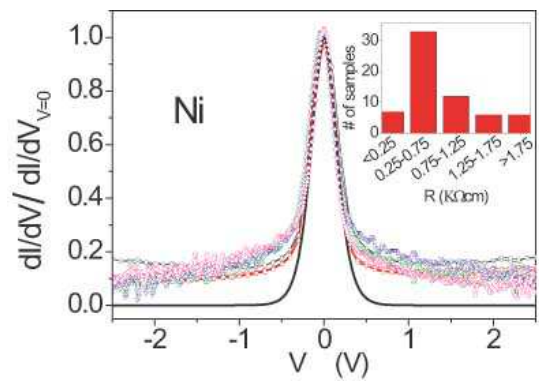}
\noindent{\caption{Normalized differential conductance for 8
different FETs with Ni electrode (open symbols) exhibiting
reproducible behavior. The continuous line is a plot of the
differential conductance of two Schottky diodes in series obtained
from Eq.2. Inset: histogram showing the spread in contact
resistance for Ni electrodes.} \label{Fig3}}
\end{figure}
The situation is very different for Ni-, Co-, and Cu-contacted
transistors. Fig. 3 and 4 show the differential conductance curves
for many FETs with Nickel and Copper electrodes, normalized to the
zero bias value. Remarkably, all curves fall nearly on top of each
other, indicating a good reproducibility of the bias dependence of
the contact resistance. The histograms shown in the insets of Fig. 3
and 4 quantify the reproducibility in the absolute (zero-bias) value
of the contact resistance. For Nickel the absolute value of the
contact resistance exhibit a good reproducibility: for the majority
of devices $dV/dI(V=0) = 500 \pm 250$ $\Omega$cm; for the few
devices for which $dV/dI(V=0) > 1000$ $\Omega$cm the larger
resistance is likely to originate from an imperfect lift-off process
during the electrode fabrication, causing a poor adhesion of the
crystal to the metal surface. For Copper -and similarly for Cobalt
(data not shown)-, the spread in contact resistance is between one
and two orders of magnitude (still considerably smaller than for
Gold). This larger spread makes the reproducibility of the bias
dependence in devices with Co and Cu contacts even more surprising.
Overall, it is a remarkable finding that materials such as Ni, Co,
and Cu, whose surface oxidizes during fabrication, lead to an enhanced
reproducibility.\\
We now take a first step in interpreting the experimental data, by
modeling our devices as two oppositely biased Schottky diodes
connected in series (corresponding to the metal/organic and
organic/metal interfaces at the source and drain contacts). The
simplest expression for the current through a diode \cite{Sze}
reads:
\begin{equation}
I(V)= I_0(e^{\frac{eV}{nkT}}-1)
\end{equation}
where $n$ is the so-called ideality factor and $I_0$ is taken to be
constant (i.e., independent of $V$). The resulting $I-V$ curve for
two back-to-back diodes $I-V$ curve then is:
\begin{equation}
I(V)= I_0 \tanh(\frac{eV}{2nkT}).
\end{equation}
The continuous lines in Fig. 3 and 4 are plots of the differential
conductance calculated by differentiating this equation, from
which we see a qualitative agreement between the data and the
back-to-back Schottky diode picture. The reproducibility of the
data is sufficient to discriminate quantitatively between the
behavior of Nickel, for which the width of the peak corresponds to
an ideality factor of $n\simeq 3$, and for Copper, where $n \simeq
1$. The deviation at high bias -i.e., the fact that the measured
differential conductance is higher than what is expected-
originates from having assumed that $I_0$ is a constant, whereas
in reality $I_0$ depends on bias. Within conventional models of
metal/semiconductor interfaces the bias dependence of $I_0$ may
originate from different microscopic phenomena, such as a
bias-induced lowering of the Schottky barrier (i.e. the Schottky
effect), diffusion limited transport, tunneling, etc. \cite{Sze}.
The reproducibility of the bias-dependent is therefore crucial to
determine which of these microscopic phenomena dominates the
behavior of charge injection at metal/organic
interfaces.\\
\begin{figure}[h]
\centering
\includegraphics[width=0.8\columnwidth]{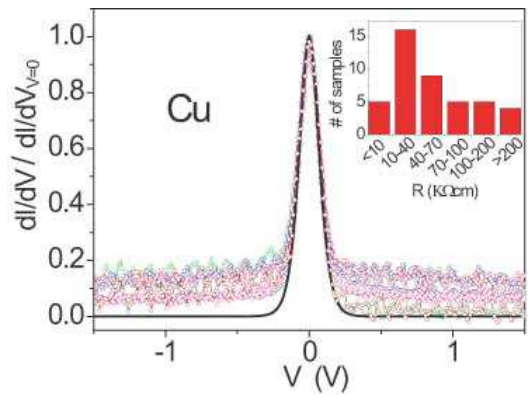}
\noindent{\caption{Normalized differential conductance for 10
different FETs with Cu electrode (open symbols) exhibiting
reproducible behavior. The continuous line is a plot of the
differential conductance of two Schottky diodes in series obtained
from Eq.2. Inset: histogram showing the spread in contact
resistance for Cu electrodes.} \label{Fig4}}
\end{figure}
In conclusion, we have investigated the contact resistance of
organic transistors, with electrodes made of five different
metals, including materials such as Cu and Co that had not been
previously used. Our results show that Ni, Cu, and Co enable a
superior reproducibility of the contact characteristics as
compared to metals used in the past. This good reproducibility
suggests the possibility to use FETs contacted with these metals
for the study of microscopic
processes of carrier injection into
 organic semiconductors.\\
Financial support from FOM, NanoNed, and NWO (Vernieuwingsimpuls)
is gratefully acknowledged.


\begin{thebibliography}{10}
\bibitem{Measurements}
Measurements were performed at room temperature, in vacuum (10
$^{-6}$ mbar) and in the dark, using either an Agilent E5270A or a
HP4192A parameter analyzer.
\end{thebibliography}
\end{document}